\documentclass[aps,prb,twocolumn,showpacs]{revtex4}
\usepackage{graphicx}

\begin{document}

\title{Scaling of Hall Resistivity in the Mixed State of MgB$_2$ Films}

\author{Huan Yang$^1$, Ying Jia$^1$, Lei Shan$^1$, Chenggang Zhuang$^{2,3,4}$,
X. X. Xi$^{3,4}$, Qi Li$^3$, Zikui Liu$^4$, Qingrong Feng$^{2}$, and
Hai-Hu Wen$^1$} \email{hhwen@aphy.iphy.ac.cn}

\address{$^1$National
Laboratory for Superconductivity, Institute of Physics and National
Laboratory for Condensed Matter Physics, Chinese Academy of
Sciences, P.O.~Box 603, Beijing 100080, P.~R.~China}
\address{$^2$Department of Physics, Peking University, Beijing
100871, P.~R.~China}
\address{$^3$Department of Physics, The Pennsylvania State
University, University Park, Pennsylvania 16802, USA}
\address{$^4$Department of Materials Science and Engineering,
The Pennsylvania State University, University Park, Pennsylvania
16802, USA}

\begin{abstract}
The longitudinal resistivity ($\rho_{xx}$) and transverse
resistivity ($\rho_{xy}$) of MgB$_2$ thin films in the mixed state
were studied in detail. We found that the temperature dependencies
of $\rho_{xx}$ and $\rho_{xy}$ at a fixed magnetic field ($H$)
satisfy the scaling law of $\rho_{xy}=A\rho_{xx}^\beta$, where the
exponent $\beta$ varies around 2.0 for different fields. In the low
field region (below 1$\;$T), $\beta$ maintains a constant value of
2.0 due to the weak pinning strength of the vortices, mainly from
the superfluid of the $\pi$ band. When $H>1\;$T, $\beta$ drops
abruptly to its lowest value at about 2$\;$T because of the
proliferation of quasiparticles from the $\pi$-band and, hence, the
motion of the vortices from the superfluid of the $\sigma$-band
dominates the dissipation. As the field is increased further, the
vortex pinning strength is weakened and $\beta$ increases
monotonically towards 2.0 at a high field. All the results presented
here are in good agreement with the expectation of the vortex
physics of a multi-band superconductor.
\end{abstract}
\pacs{74.70.Ad, 74.25.Qt, 74.25.Fy}

\maketitle

\section{Introduction}
Since the discovery of the binary superconductor MgB$_2$ in 2001, a
number of studies have reported that the properties of its mixed
state
\cite{coherence_length,MgB2Wen,JiaY,twogap,MgB2IV,tunnelling,STMvortex}
and normal state\cite{MR_Pallecchi,LiQ,Mazin} properties have a
close relationship with its multiband characteristics. The two
three-dimension (3D) $\pi$ bands and two quasi-two-dimension
(quasi-2D) $\sigma$ bands provide rich physics compared with those
materials with a single band. \cite{review} The two sets of bands
have different superconducting gaps, i.e., about 7$\;$meV for the
$\sigma$ bands, and about 2$\;$meV for the $\pi$
bands.\cite{twogap,gaps} Moreover, the coherence length of the $\pi$
bands is much larger than that of the $\sigma$ bands
\cite{coherence_length}. Many experiments have demonstrated that the
$\pi$-band pairing strength is closely related to that of the
$\sigma$-band and both the interband and intraband scattering play
important roles in this multiband system. Owing to the complicated
nature of superconductivity in this system, its vortex dynamics may
exhibit some interesting or novel features.

For the high critical temperature superconductors (HTSC), the Hall
measurements revealed a puzzling scaling relationship between
temperature dependence $\rho_{xx}$ and $\rho_{xy}$ in the mixed
state at a fixed field, i.e., $\rho_{xy}=A\rho_{xx}^\beta$, where
the exponent $\beta$ was observed in the range from 1.5 to 2.0 in
YBaCuO \cite{HallYBCO}, YCaBaCuO \cite{WangYCBCO} and HgBaCaCuO
\cite{HallHg1212}, $\beta\sim2$ in BiSrCaCuO \cite{HallBi2212} and
TlBaCaCuO \cite{HallTl2212}. A number of theories have been proposed
in order to explain this scaling law. First, $\beta$ may be relevant
to the scaling parameters of $I-V$ curves near the vortex-glass
transition temperature. \cite{Dorsey} Another theoretical model
proposed by Vinokur {\it et al.} \cite{ThVinokur} suggests that the
scaling exponent $\beta$ should be 2.0 independent of the vortex
pining strength. Later on, Wang {\it et al.} \cite{ThWang} developed
 a theory considering both the pinning effect and the thermal
fluctuations, and gave the range of $\beta$ from 1.5 to 2.0
depending on the strength of the vortex pinning. For HTSCs, the
experimental results revealed a universal $\beta$ value about 2.0 at
a higher field and a small value at a lower field.

The Hall effect in MgB$_2$ has been studied in many previous works,
and the anomalous Hall effect was found \cite{UnusualHallMgB2}. Jung
{\it et al.} proposed that the field dependence of Hall conductivity
$\sigma_{xy}$ may result from both the vortex motion and the
quasiparticles. \cite{Hallsigma} Similar to the HTSCs, a universal
scaling behavior of the Hall resistivity was found in the films,
which gives a constant value of $\beta$ of $2.0\pm0.1$ independent
of the magnetic field, the temperature, and the current density,
even for the behavior of the field dependence of the two
resistivities. \cite{MgB2Scaling} In this paper, we present the Hall
scaling result in the mixed state, and our results are closely
related to the multiband property.

\section{Experiments}

The MgB$_2$ thin films used in this work were grown by the hybrid
physical-chemical vapor deposition technique \cite{film} on (0001)
6$H$-SiC substrates. The films were epitaxial and highly $c$-axis
orientated with a thickness of about 100$\;$nm. There are no
impurity peaks in the X-ray diffraction pattern, and the sharp
intrinsic peaks show good crystallinity; the $\phi$ scan (azimuthal
scan) indicates the sixfold hexagonal symmetry of the MgB$_2$ film
matching the substrate. The scanning electron microscopy (SEM) image
shows no observable grain boundaries in micrometer scale, suggesting
a good homogeneity of the films. All samples have a critical
temperature of about 40$\;$K with a very sharp resistive transition.
A thin layer of gold was sputtered onto the electrode parts of the
film, and the contact resistance is smaller than 1$\;\Omega$. In the
measurement, bidirectional current mode was applied with a current
density of $10^3\;\mathrm{A/cm}^2$ in the linear resistivity region
\cite{MgB2IV}. For small current and contact resistance, the
self-heating effect of the current in the measurement can be
omitted. The temperature dependencies of $\rho_{xx}$ and $\rho_{xy}$
were measured at the same time at various magnetic fields applied
perpendicular to the $ab$ plane of the films, and the voltage
resolution was about $10\;$nV. The Hall resistivity $\rho_{xy}$ was
obtained by averaging the results measured in the inverted fields.

\section{Results and discussion}

\begin{figure}
\includegraphics[width=8cm]{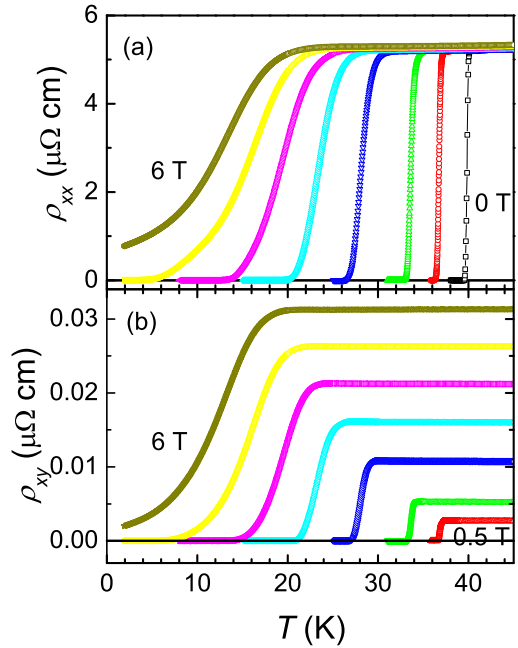}
\caption{(color online) Temperature dependence of longitudinal
resistivity $\rho_{xx}$ (a) and transverse resistivity $\rho_{xy}$
(b). In (a), from right to left, the corresponding fields are: 0,
0.5, 1, 2, 3, 4, 5, 6 T; in (b), from right to left, the
corresponding fields are: 0.5, 1, 2, 3, 4, 5 and 6 T. } \label{fig1}
\end{figure}

Fig.~\ref{fig1} shows the temperature dependence of $\rho_{xx}$ and
$\rho_{xy}$ of a MgB$_2$ film measured in various magnetic fields.
The transition width of $\rho_{xx}$ at 0$\;$T is about 0.2$\;$K, and
the residual resistivity $\rho_\mathrm{n}$ at 42$\;$K is about
5.3$\;\mu\Omega\,$cm. The residual resistance ratio
$RRR\equiv\rho(300\;\mathrm{K})/\rho(42\;\mathrm{K})$ is about 5.1.
There is a continuous broadening of transition width as the field
increases. At a field larger than 5$\;$T, there is a nonvanishing
dissipation at the lowest temperature in the measurement, i.e.,
1.9$\;$K, and it has been discussed in detail elsewhere
\cite{MgB2Wen,JiaY}.

\begin{figure}
\includegraphics[width=8cm]{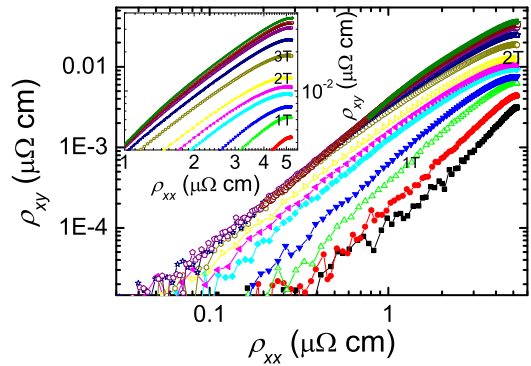}
\caption{(color online)Correlation between $\rho_{xx}$ and
$\rho_{xy}$ measured at magnetic fields of 0.5, 0.7, 1.0, 1.2, 1.5,
1.7, 2.0, 3.0, 4.0, 5.0, 5.5 and 6 T (from bottom to top) in log-log
plot. The inset shows the high resistivity section. } \label{fig2}
\end{figure}

\begin{figure}
\includegraphics[width=8cm]{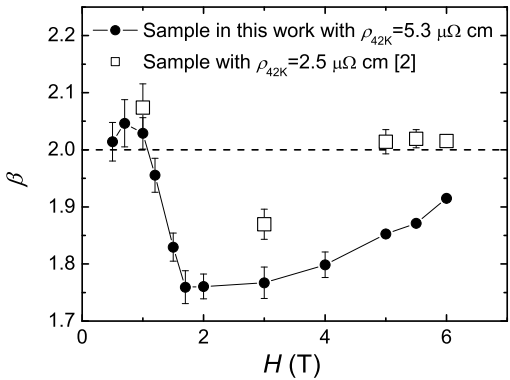}
\caption{Magnetic field dependence of the exponent $\beta$ value by
the linear fit to the log-log plot of Hall and transverse
resistivities in the small resistivity area of this work (solid
circles) and from our previous work \cite{JiaY} (open squares). }
\label{fig3}
\end{figure}

\begin{figure}
\includegraphics[width=8cm]{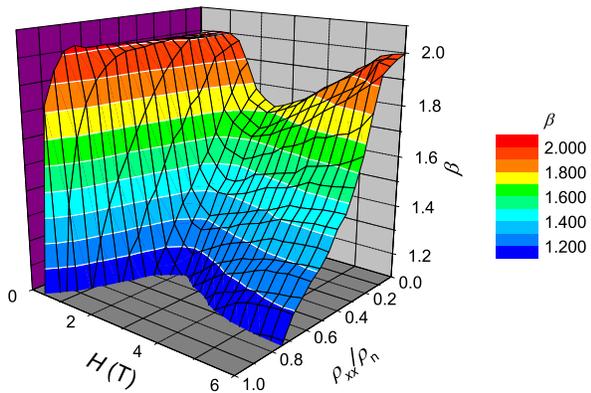}
\caption{(color online) 3D plot of the local differential $\beta$
versus the magnetic fields and normalized longitudinal resistivity.
In the region with low resisitivity, $\beta$ locates between
$1.7-2.0$ depending on the field, indicating the vortex motion. At
high resistivity region, $\beta$ drops rapidly, showing the
contribution by both the vortices and quasiparticles.} \label{fig4}
\end{figure}

In Fig.~\ref{fig2}, we plot $\rho_{xy}$ vs $\rho_{xx}$ at different
fields from 0.5 to 6$\;$T in double logarithmic scales. Linear
behavior can be seen clearly in the region of
$\rho_{xx}<1/4\rho_\mathrm{n}$ for almost all fields.
Fig.~\ref{fig3} shows the field dependence of the slope $\beta$
determined by linearly fitting the small resistance regime presented
in Fig.~\ref{fig2}. As the field increases, $\beta$ remains constant
at 2.0 below 1$\;$T, drops rapidly to a minimal value around 2$\;$T,
and increases monotonically at higher fields. Similar results were
observed for all studied samples. As an example, the $\beta$ values
obtained from our previous work \cite{JiaY} are also plotted in the
figure with open squares. Although the sample studied in the
previous work has a lower residual resistivity, the determined slope
shows a minimum at an intermediate field of 3$\;$T, which is in good
agreement with the data obtained in this work. In Fig.~\ref{fig4},
we plot the smoothed differential value of $\beta$ versus the
magnetic field and the normalized resistivity in a 3D plot, and the
white line in the surface is the state with the same $\beta$. This
shows that, when $\rho_{xx}/\rho_\mathrm{n}<20\%$, although there is
a minimum value at about 2$\;$T, $\beta$ locates between
$1.7\sim2.0$ depending on the magnetic field similar to the HTSC,
indicating the vortex motion. Surprisingly, the value reaches 2.0
and the curves overlap at the high field, starting from 3$\;$T, and
the low resistivity region. While, when $\rho_{xx}/\rho_\mathrm{n}$
increases, the $\beta$ value drops rapidly, showing the joint
contribution by both the vortices and the quasiparticles.

In the mixed state of a type II superconductor, the longitudinal and
Hall resistivity can be related by the following scaling
law\cite{ThVinokur}:
\begin{equation}
\rho_{xy}=\frac{c^2\alpha}{\Phi_0B}\rho_{xx}^2. \label{eq1}
\end{equation}
which is appropriate for any type of vortex motion including flux
flow, thermally assisted flux flow (TAFF), and vortex creep in the
vortex glass regime. Wang {\it et al.} proposed another form
expressed\cite{ThWang} by:
\begin{equation}
\rho_{xy}=\frac{\beta_0\rho_{xx}^2}{\Phi_0B}\left\{\eta(1-\overline{\gamma})-2{\overline{\gamma}}\Gamma(v_L)\right\}.
\label{eq2}
\end{equation}
in which $\beta_0=\mu_mH_{c2}$, where $\mu_m$ is the mobility of the
charge carrier and $H_{c2}$ is the usual upper critical field;
$\Phi_0$ is the quantum of the flux;
$\overline{\gamma}=\gamma(1-\overline{H}/H_{c2})$ with
$\overline{H}$ the average magnetic field over the core and $\gamma$
the parameter describing contact force on the surface of core.
$\Gamma(v_L)$ is a positive scale function dependent on the
time-average flux-motion velocity $v_L$ and the Lorentz force. For
$\xi/l\ll1$ with $\xi$ the coherence length and $l$ the mean free
path of the charge carrier, $\gamma\sim0$ (Nozi\`{e}res-Vinen
limit), and for $\xi/l\ge1$, $\gamma\sim1$ (Bardeen-Stephen limit).
For MgB$_2$, the coherence length of the $\pi$ band is about
50$\;$nm as derived from the STM measurement \cite{STMvortex}, and
the calculated value from the upper critical field of the $\sigma$
band is about $20\;$nm. The mean free path of this film is about
25$\;$nm estimated from the resistivity measurement. Hence, the
value of $\gamma$ can be determined to be about 1. In a weak pinning
system, $\Gamma(v_L)\ll\eta\overline{H}/H_\mathrm{c2}$, so
Eq.~\ref{eq2} becomes $\rho_{xy}\sim A\rho_{xx}^2$ which is similar
to Eq.~\ref{eq1} with $A$ approximately being a field independent
constant. However, for the strong pinning case
$\Gamma(v_L)\gg\eta\overline{H}/H_\mathrm{c2}$, and with the rough
estimation $\Gamma(v_L)\sim
v_L^{-1/2}\sim\rho_{xx}^{-1/2}$,\cite{SPVinokur} the scaling
behavior changes to $\rho_{xy}\sim A\rho_{xx}^{1.5}$. With the
decrease of pinning strength, $\beta$ changes continuously from 1.5
to 2.0. As shown in Fig.~\ref{fig3}, the continuous increase of
$\beta$ above 2$\;$T indicates that the pinning strength dominates
in the scaling law. Moreover, the $\beta$ value of the cleaner film
used in previous work ($\rho(42K)=2.45\;\mu\Omega\,$cm) is clearly
larger than that of the the dirtier sample here
($\rho(42K)=5.3\;\mu\Omega\,$cm), which also suggests the dominant
role of the pinning strength in the current system. For the clean
films, at a field above 4$\;$T, $\beta$ remains at 2.0 with an
almost constant coefficient $A$. Such a weak pinning regime in the
low temperature region at a higher field suggests that the vortex
motion associated with the vortex quantum fluctuation is a possible
origin of the non-vanishing dissipation in higher fields below
$H_\mathrm{c2}$, as addressed in detail elsewhere. \cite{JiaY}

The scaling law of the MgB$_2$ system observed here seems to be
similar to the situation of HTSC at high fields. However, at a field
lower than $1\;$T, $\beta$ gives a puzzling value of 2.0 independent
of the magnetic field. As proposed by Dorsey {\it et al.}
\cite{Dorsey}, $\beta=1+\lambda_v/(z+2-D)$, where $\lambda_v$ is an
eigenvalue, while $z$ and $D$ are the parameters in the $I-V$
vortex-glass scaling. In our previous work \cite{MgB2IV}, we gave
the value of $z$ and $D$ of the cleaner film mentioned above. The
obtained $z$ value at $3\;$T is larger than that of small fields
from 0.1 to 1$\;$T; while the $v$ value has an opposite field
dependence, which is consistent with the low-field behavior of
$\beta$ presented in this work. All these aspects indicate that the
vortex state at a field lower than $1\;$T is abnormal in MgB$_2$
compared with a single band superconductor.

MgB$_2$ is a two-band superconductor, and the $\pi$-band
superconductivity can be easily destroyed by a small magnetic field.
From the theoretical calculation \cite{QPTh} and the point contact
spectroscopy,\cite{QPtunneling} we find that at the magnetic field
around $1\;$T there is an inflexion in the plot of the density of
quasiparticles versus magnetic field. In other words, at small
fields below $1\;$T, the $\pi$-band pairing is depressed rapidly and
enormous quasiparticles are generated. In this regime, the vortices
from the superfluid dominated by the $\pi$ band may play an
important role in the conducting property, and the spread of the
quasiparticles both outside and inside the vortex cores
\cite{STMvortex} could reduce the pinning strength of the $\pi$-band
dominated vortices. At fields higher than $1\;$T, the pairing
strength of the $\pi$ band is sustained by the coupling to the
$\sigma$ band and the proliferation of quasiparticles increases
slowly. Consequently, the $\sigma$-band contribution to the vortex
behavior becomes increasingly important. The vortices from the
superfluid dominated by the $\sigma$ band may be easily pinned and,
hence, the exponent $\beta$ drops rapidly to a small value. Finally,
when the field increases higher than $2\;$T, the $\sigma$-band
dominated vortices play an important role in the transport property,
and the situation is very similar to that of HTSC. Furthermore, from
the experiment of small-angel neutron scattering
\cite{neutronVortex}, an obvious vortex structure change from 0.5 to
$0.9\;$T was observed in the single crystal sample, which may imply
that the $\pi$-band vortex lattice changes to the $\sigma$-band
vortex lattice, though this phenomenon still needs further
consideration.

\section{Conclusions}

We have presented the scaling law between the Hall and longitudinal
resistivity at fixed magnetic fields for samples with different
pinning strengths. The scaling exponent $\beta$ equals 2.0 at low
fields (below $1\;$T), which may be associated with the weak-pinning
$\pi$-band dominated vortices. When the field reaches $2\;$T, the
$\sigma$-band dominated vortices play an important role in the
conductivity, and the stronger pinning strength reduces the value of
$\beta$. By further increasing the field, the vortex pinning is
weakened continuously and $\beta$ increases and finally approaches
2.0 in the low-resistivity region again. This is another proof that
the multiband property is very important in the mixed state of
MgB$_2$.

\section*{Acknowledgments}
This work is supported by the National Science Foundation of China,
the Ministry of Science and Technology of China (973 project:
2006CB601000 and 2006CB921802), and the Knowledge Innovation Project
of the Chinese Academy of Sciences (ITSNEM). The work at Penn State
is supported by NSF under Grants Nos. DMR-0306746 (X.X.X.),
DMR-0405502 (Q.L.), and DMR-0514592 (Z.K.L. and X.X.X.), and by ONR
under grant No. N00014-00-1-0294 (X.X.X.).


\begin{thebibliography}{10}
\bibitem{coherence_length} A. E. Koshelev, and A. A. Golubov, Phys. Rev. Lett.
 {\bf92} 107008 (2004); A. A. Golubov, and A. E. Koshelev, Phys. Rev. B {\bf68} 104503 (2003).

\bibitem{MgB2Wen} H. H. Wen, S. L. Li, Z. W. Zhao, Y. M. Ni,
Z. A. Ren, G. C. Che, H. P. Yang, Z. Y. Liu, and Z. X. Zhao, Chin.
Phys. Lett. {\bf 18}, 816 (2001); H. H. Wen, S. L. Li, Z. W. Zhao,
H. Jin, Y. M. Ni, W. N. Kang, H. J. Kim, E. M. Choi, and S. I. Lee
{\it Phys. Rev. B} {\bf 64}, 134505 (2001).

\bibitem{JiaY} Y. Jia, H. Yang, Y. Huang, L. Shan, C. Ren, C. G. Zhuang, Y. Cui, Q. Li,
Z. K. Liu, X. X. Xi, and H. H. Wen, arXiv:cond-mat/0703637
(unpubilished)  (2007).

\bibitem{twogap} H. J. Choi, D. Roundy, H. Sun, M. L. Cohen, and S. G.
Louie, Nature (London) {\bf418}, 758 (2002).

\bibitem{MgB2IV} H. Yang, Y. Jia, L. Shan, Y. Z. Zhang, H. H. Wen, C. G. Zhuang, Z. K. Liu,
Q. Li, Y. Cui, and X. X. Xi, Phys. Rev. B {\bf76}, 134513 (2007).

\bibitem{tunnelling} R. S. Gonnelli, D. Daghero, G. A. Ummarino, V. A. Stepanov, J. Jun,
 S. M. Kazakov, and J. Karpinski, Phys. Rev. Lett. {\bf89}, 247004 (2002).

\bibitem{STMvortex}M. R. Eskildsen, M. Kugler, S. Tanaka, J. Jun, S. M. Kazakov,
 J. Karpinski, and {\O}. Fischer, Phys. Rev. Lett. {\bf89}, 187003 (2002).

\bibitem{MR_Pallecchi} I. Pallecchi, V. Ferrando, E. Galleani D'Agliano,
 D. Marr\'{e}, M. Monni, M. Putti, C. Tarantini, F. Gatti, H. U. Aebersold, E. Lehmann,
 X. X. Xi, E. G. Haanappel, and C. Ferdeghini, Phys. Rev. B {\bf72},
 184512 (2005).

\bibitem{LiQ} Qi Li, B. T. Liu, Y. F. Hu, J. Chen, H. Gao, L. Shan, H. H. Wen, A. V. Pogrebnyakov,
J. M. Redwing, and X. X. Xi, Phys. Rev. Lett. {\bf96}, 167003
(2006).

\bibitem{Mazin} I. I. Mazin, O. K. Andersen, O. Jepsen, O. V. Dolgov, J. Kortus, A. A. Golubov,
 A. B. Kuz'menko, and D. van der Marel, Phys. Rev. Lett. {\bf89},
 107002 (2002).

\bibitem{review} P. C. Canfield, and G. Crabtree, Phys. Today {\bf56},
34 (2003).

\bibitem{gaps} M. Iavarone, G. Karapetrov, A. E. Koshelev, W. K. Kwok,
G. W. Crabtree, D. G. Hinks, W. N. Kang, E. M. Choi, H. J. Kim, H.
J. Kim, and S. I. Lee, Phys. Rev. Lett. {\bf89}, 187002 (2002).

\bibitem{HallYBCO} J. Luo, T. P. Orlando, J. M. Graybeal, X. D. Wu, and R. Muenchausen
Phys. Rev. Lett. {\bf 68}, 690 (1992); W. N. Kang, D. H. Kim, S. Y.
Shim, J. H. Park, T. S. Hahn, S. S. Choi, W. C. Lee, J. D.
Hettinger, K. E. Gray, and B. Glagola, Phys. Rev. Lett. {\bf76},
2993 (1996).

\bibitem{WangYCBCO} Z. Wang, Y. Z. Zhang, X. F. Lu, H. Gao, L. Shan, and H. H.
Wen, Physica C {\bf 422}, 41 (2005).

\bibitem{HallHg1212} W. N. Kang, S. H. Yun, J. Z. Wu, and D. H. Kim, Phys. Rev. B {\bf
55}, 621 (1997).

\bibitem{HallBi2212} A. V. Samoilov, Phys. Rev. Lett. {\bf71},
617 (1993).

\bibitem{HallTl2212} A. V. Samoilov, Z. G. Ivanov, and L. G. Johansson, Phys. Rev. B {\bf
49}, 3667 (1994).

\bibitem{Dorsey} A. T. Dorsey, and M. P. A. Fisher, Phys. Rev. Lett. {\bf
68}, 694 (1992).

\bibitem{ThVinokur} V. M. Vinokur, V. B. Geshkenbein, M. V. Feigel'man, and
G. Blatter, Phys. Rev. Lett. {\bf71}, 1242 (1993).

\bibitem{ThWang} Z. D. Wang, J. M. Dong, and C. S Ting, Phys. Rev. Lett.
{\bf72}, 3875 (1994).

\bibitem{UnusualHallMgB2} R. Jin, M. Paranthaman, H. Y. Zhai, H. M. Christen,
D. K. Christen, and D. Mandrus, Phys. Rev. B {\bf 64}, 220506(R)
(2001).

\bibitem{Hallsigma} S. G. Jung, W. K. Seong, J. Y. Huh, T. G. Lee, W. N. Kang,
E. M. Choi, H. J. Kim, and S. I. Lee, Supercond. Sci. Technol. {\bf
20}, 129 (2007).

\bibitem{MgB2Scaling} W. N. Kang, H. J. Kim, E. M. Choi, H. J. Kim, K. H. P.
Kim, and S. I. Lee, Phys. Rev. B {\bf 65}, 184520 (2002).

\bibitem{film} X. H. Zeng, A. J. Pogrebnyakov, A. Kotcharov, J. E.
Jones, X. X. Xi, E. M. Lysczek, J. M. Redwing, S. Y. Xu, Q. Li, J.
Lettieri, D. G. Schlom, W. Tian, X. Q. Pan, and Z. K. Liu, Nature
Mater. {\bf1}, 1 (2002).

\bibitem{SPVinokur} V. M. Vinokur, V. B. Geshkenbein, M. V. Feigel¡¯man, and G.
Blatter, Phys. Rev. Lett. {\bf 71}, 1242 (1993).

\bibitem{QPTh} A. E. Koshelev, and A. A. Golubov, Phys. Rev. Lett. {\bf
90}, 177002 (2003).

\bibitem{QPtunneling} R. S. Gonnelli, D. Daghero, A. Calzolari, G. A. Ummarino,
V. Dellarocca, V. A. Stepanov, J. Jun, S. M. Kazakov, and J.
Karpinski, Phys. Rev. B {\bf 69}, 100504(R) (2004); Y. Bugoslavsky,
Y. Miyoshi, G. K. Perkins, A. D. Caplin, L. F. Cohen, A. V.
Pogrebnyakov, and X. X. Xi, Phys. Rev. B {\bf 72} 224506 (2005); A.
Kohen, F. Giubileo, Th. Proslier, F. Bobba, A. M. Cucolo, W. Sacks,
Y. Noat, A. Troianovski, and D. Roditchev, Eur. Phys. J. B {\bf 57},
21 (2007).

\bibitem{neutronVortex} R. Cubitt, M. R. Eskildsen, C. D. Dewhurst, J. Jun,
S. M. Kazakov, J. Karpinski, Phys. Rev. Lett. {\bf 91}, 047002
(2003).

\end{thebibliography}
\end{document}